\documentclass[preprint,epsfig,onecolumn,floats,showpacs]{revtex4}
\usepackage{amssymb}
\usepackage{graphicx}
\usepackage{epsfig}
\usepackage{amsmath}
\usepackage{amsfonts}
\usepackage{bm}
\usepackage{subfigure}

\begin{document}

\title{Quantum Bus of Metal Nanowire with Surface Plasmon Polaritons}
\author{Guo-Ping Guo}
\email{gpguo@ustc.edu.cn}
\author{Zhi-Rong Lin}
\author{Tao Tu}
\email{tutao@ustc.edu.cn}
\author{Hai-Ou Li}
\author{Chang-Ling Zou}
\author{Xi-Feng Ren}
\author{Guang-Can Guo}
\affiliation{Key Laboratory of Quantum Information, University of Science and Technology
of China, Chinese Academy of Sciences, Hefei 230026, P. R. China}
\date{\today }

\begin{abstract}
We develop an architecture for distributed quantum computation using quantum
bus of plasmonic circuits and spin qubits in self-assembled quantum dots.
Deterministic quantum gates between two distant spin qubits can be reached
by using an adiabatic approach in which quantum dots couple with highly
detuned plasmon modes in a metallic nanowire. Plasmonic quantum bus offers a
robust and scalable platform for quantum optics experiments and the
development of on-chip quantum networks composed of various quantum nodes,
such as quantum dots, molecules and nanoparticles.
\end{abstract}

\maketitle

\baselineskip20pt%

Plasmonic circuits, providing the ability to integrate electronics and
optics on the nanoscale, may lead an exciting application to carry classical
information between microprocessors in integrated chip \cite{Ozbay2006}. For
a scalable quantum chip, it is also desired to find a high speed on-chip
quantum bus where the information can be coherently transferred between
distant processing nodes. Integrated plasmonic circuits are an attractive
route towards realizing such promise since they allow for scalability and
coherent coupling to single emitters \cite%
{Lukin2006PRL,Lukin2007PRB,Lukin2007Nature,Lukin2007NP}. Recently,
self-assembled quantum dots are argued as a promising candidate for building
a practical quantum processor for their potential advantages, including
self-evident scaling, ultra-fast coherent control and long lived spin states
\cite{Calarco,Nazir,Nazir2008,Machnikowshi}. Picosecond optical coherence
measurement, preparation and manipulation of electron spin states have been
demonstrated in self-assembled quantum dot systems \cite%
{Awschalom2006,Imamoglu2007,Awschalom2007}. However, most of the proposals
for coherent coupling of two spin qubits are based on the interactions
between neighboring dots \cite{Sandoghdar2002,Imamoglu2008}. A solution is
expected to utilize quantum bus to couple qubits in a non-local and
switchable way. Here we show the implementation of a quantum bus, using the
surface plasmon polaritons confined in the metallic nanowires, to coherently
couple an arbitrary pair of distant semiconductor quantum dot spin qubits.
The interaction is mediated by the exchange of virtual surface plasmon
polaritons rather than real ones, avoiding the decoherence of the system.
Using adiabatic control of the qubits, we demonstrate high fidelity quantum
operations between spatially separated spin qubits. Our approach is
applicable to a wide class of electronic spin qubits near the nanowire and
can be used for the implementation of distributed quantum computing
architectures.

Fig. 1 shows our scalable solid state quantum computer architecture, in
which spins in self-assembled quantum dots (QDs) act as qubits. These QDs
can be grown by molecular beam epitaxy along the $z$ axis. Single electron
can be deterministically injected to QD by the semi-transparent metallic
gate. Each QD is selectively coupled to a laser field guided by a nanotip
and a fiber taper which has been realized in recent experiment \cite{Ren2009}%
. The coupling between QDs is mediated by surface plasmon modes in metallic
nanowire. Throughout, we use a silver nanowire with electric permittivity $%
\epsilon _{2}=-50+0.6i$ at room temperature and a vacuum wavelength $\lambda
_{0}=950$ nm \cite{Kroner}, and the surrounding dielectric $\epsilon _{1}=2$
(the nanowire is covered with a thin layer of PMMA). We assume a large 
oscillator strength QDs $f=100$ which corresponds lateral radius of around 
22nm \cite{Peter}. In the following, we discuss the quantization of surface 
plasmons in metal nanostructure and then give the detailed coupling mechanism 
between QD spin and plasmon.

\begin{figure}[tbp]
\includegraphics[width=0.8\columnwidth]{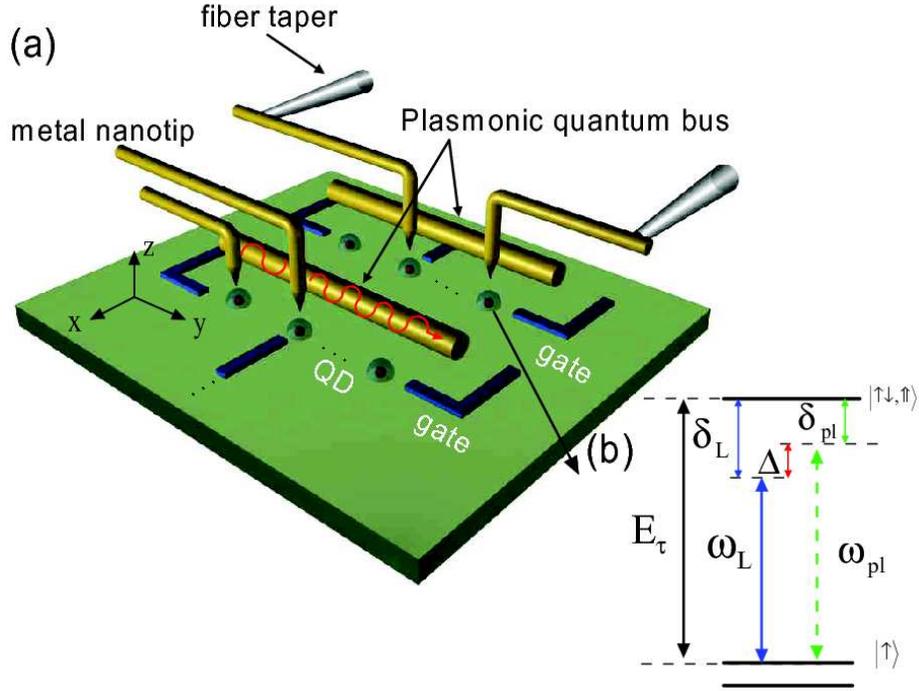}
\caption{(a) The hybrid platform to implement quantum information
processing. In this architecture, non-local spins in self-assembled quantum
dots (QDs) are interconnected via quantum bus of surface plasmons (red wave
line) in metal nanowire. Single electron can be deterministically injected
to QD by static gate. The qubit is individually addressed using a laser
guided by a nanotip and a fiber taper. (b) The levels of a charged quantum
dot. The spin state $|\uparrow \rangle $ is coupled to an optically excited
trion state $|\uparrow \downarrow, \Uparrow\rangle $ by plasmon modes and
laser pulse.}
\end{figure}

Surface plasmons, or surface plasmon polaritons are generally treated as
classical electromagnetic waves that propagate along surface of a conductor.
Recent experiment has shown that the fluorescence of quantum dots can be
coupled to metallic nanowires and single, quantized plasmon can be generated
\cite{Lukin2007Nature}. In this work, we present a fully quantum mechanical
approach for the interaction between electron spin in QD and quantized
plasmon in the nanowire. In a cylindrical nanowire, the solution to
electromagnetic modes has been given in Ref. \cite{Arista2001,Arista2003}
for quite some time. The energy of the surface plasmon field is the sum of
the kinetic and electrostatic energy
\begin{equation}
H=\frac{1}{2}\int \{n_{0}m_{e}(\nabla \Psi _{s})^{2}+\rho _{s}\Phi
_{s}\}d^{3}r,  \label{PlHamiltonian}
\end{equation}%
where $n_{0}$ is the equilibrium electron density in the metal, $m_{e}$ is
the electron mass, $\Psi _{s}(\bm r)$ is the velocity potential, $\rho _{s}(%
\bm r)$ is the charge density displacement from equilibrium, and $\Phi _{s}(%
\bm r)$ is the scalar potential at the surface. The velocity potential $\Psi
_{s}(\bm r)$ and electronic density displacement $n_{s}(\bm r)$ satisfies
the continuity equation
\begin{equation}
\partial _{t}n_{s}=n_{0}\nabla ^{2}\Psi _{s}.
\end{equation}%
For a cylindrical nanowire of radius $R$ and length $L$, the velocity
potential and electrical potential can be expanded in cylindrical
coordinates
\begin{eqnarray}
\Psi _{s}(\bm r) &=&\sum_{k,m}a_{k,m}K_{m}(k\rho )e^{i(kz+m\phi )},\Phi _{s}(%
{\bm r})=\sum_{k,m}c_{k,m}K_{m}(k\rho )e^{i(kz+m\phi )},\text{for }\rho >R,
\\
\Psi _{s}(\bm r) &=&0,\Phi _{s}({\bm r})=\sum_{k,m}b_{k,m}I_{m}(k\rho
)e^{i(kz+m\phi )},\text{for }\rho <R,
\end{eqnarray}%
and the electronic density is correspondingly expressed as
\begin{equation}
n_{s}(\bm r)=\sum_{k,m}n_{k,m}e^{i(kz+m\phi )}\delta (\rho -R),
\end{equation}%
where $I_{m}(x)$ and $K_{m}(x)$ are the cylindrical Bessel functions.
Applying Maxwell's boundary conditions, the relations between the
coefficients $a_{k,m}$, $b_{k,m}$, $c_{k,m}$ and $n_{k,m}$ can be
determined. With the electromagnetic field quantization calculations \cite%
{Scully}, the Hamiltonian of the surface plasmon field Eq. (\ref%
{PlHamiltonian}) can be expressed in the standard second quantized form
\begin{equation}
H_{pl}=\sum_{k,m}\hbar \omega _{k,m}[a_{k,m}^{\dagger }a_{k,m}+\frac{1}{2}],
\end{equation}%
where $\omega _{k,m}^{2}=\omega _{p}^{2}[\frac{kRI_{m}^{\prime }(kR)K_{m}(kR)%
}{\epsilon _{1}+(\epsilon _{s}-\epsilon _{1})kRI_{m}^{\prime }(kR)K_{m}(kR)}]
$, $\omega _{p}$ is the bulk plasma frequency, $\epsilon _{s}=3.3$ is the
background dielectric constant of the silver, and $I_{m}^{^{\prime
}}(x)=dI_{m}(x)/dx$.

We consider single-charged self-assembled QDs with strong confinement along the growth direction. Because of
the large heavy-hole-light-hole splitting, we can neglect the light-hole excitons and the QD
can be effectively described by a four-level system, the electron spin
states $|\downarrow \rangle $, $|\uparrow \rangle $, and the two trion
states consisting of two spin paired electrons and unpaired heavy hole, $%
|\uparrow \downarrow ,\Downarrow \rangle $, $|\uparrow \downarrow ,\Uparrow
\rangle $. The qubit is encoded in spin states of the excess electron. The $%
\sigma_{+}$ polarized light guided by a metallic nanotip can only connect
spin state $|\uparrow \rangle $ to an optically excited trion state $%
|\uparrow \downarrow, \Uparrow\rangle $ as shown in Fig. 1(b). Using the
rotating wave approximation, the Hamiltonian between the QD and circularly
polarized laser is expressed as
\begin{equation}
H_{laser-QD}= \delta_{L}|\uparrow \downarrow,\Uparrow \rangle \langle
\uparrow \downarrow,\Uparrow | + \Omega (t)( |\uparrow \rangle \langle
\uparrow \downarrow,\Uparrow | + h.c. ).
\end{equation}
Here $\Omega (t)$ is the optical Rabi amplitude due to laser field guided by
nanotip, $\delta _{L}=E_{\tau }-\nu $, $E_{\tau}$ is the trion energy, and $%
\nu$ is the laser frequency.

\begin{figure}[tbp]
\includegraphics[width=0.7\columnwidth]{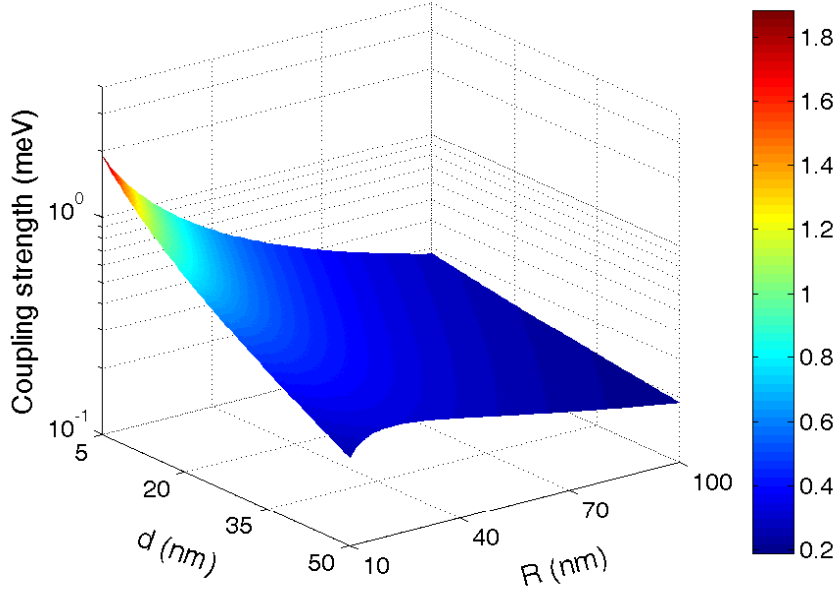}
\caption{Coupling strength $g$ as a function of $R$ and $d$ for a nanowire
length of $L = 10$ $\protect\mu$m.}
\end{figure}

The interaction between the plasmon field and a QD is given by
\begin{equation}
H_{pl-QD}=-\bm d\cdot \bm E.  \label{interaction}
\end{equation}%
where $\bm d$ is dipole operator of QD, and $\bm E$ is electric field of
surface plasmon. In the nanowire limit ($|k|R\ll 1$) all higher order modes
are cutoff, thus we only consider the fundamental mode ($m=0$), in which
longitudinal wave vector $k_{\parallel }$ propagating along nanowire
satisfies a specifical condition. The electric field component $E_{\varphi }$
($\varphi $ direction) vanishes for the fundamental mode \cite{Lukin2007PRB}%
. It's assumed that the electric field component $E_{\rho }$ (the radial
direction) has the main contribution to spin quantum number ($S=1$) of
surface plasmon \cite{Nakamura}. Thus we use $\bm E=E_{\rho }\hat{\rho}%
=-\sum_{k,m}ke^{i(kz+m\phi )}c_{k,m}K_{m}^{\prime }(k(R+d))\hat{\rho}$,
where $d$ is the distance between QD and nanowire edge, $\hat{\rho}=(-\hat{%
\epsilon ^{+}}+\hat{\epsilon ^{-}})/\sqrt{2}$, and $\hat{\epsilon ^{+}}$, $%
\hat{\epsilon ^{-}}$ represent the unit vectors of circularly polarized
light. If the longitudinal component of electric field $E_{z}$ ($z$
direction) has contribution to spin angular momentum of surface plasmon, the
following results will be obtained in a similar way. Therefore, the
interaction Hamiltonian between the plasmon field and QDs in the rotating
wave approximation in the basis $\{|\downarrow \rangle ,|\uparrow \rangle
,|\uparrow \downarrow ,\Uparrow \rangle \}$ is
\begin{equation}
H_{pl-QD}=g(a^{+}|\uparrow \rangle \langle \uparrow \downarrow ,\Uparrow
|e^{-i\delta _{pl}}+h.c.),  \label{SPP-QD}
\end{equation}%
where $g$ is the coupling strength between the QD and metal nanowire, $a^{+}$%
, $a$ are the creation and annihilation operators for fundamental surface
plasmon mode of the nanowire with a specifical wave vector $k$, the detuning
is $\delta _{pl}=E_{\tau }-\omega _{0}(k)$. From the Eq. (\ref{interaction})
and Eq. (\ref{SPP-QD}), the coupling strength between the nanowire and QD is
found to be
\begin{equation}
g=\hbar C[\frac{1}{4\pi \lbrack \epsilon _{1}+(\epsilon _{s}-\epsilon
_{1})CI_{m}^{\prime }(C)K_{m}(C)]\epsilon _{0}}\frac{\pi e^{2}f}{m_{e}LR^{2}}%
\frac{\omega _{0}}{E_{\tau }}\frac{I_{0}(C)}{K_{0}(C)}]^{1/2}K_{1}(k_{%
\parallel }(R+d)),  \label{Coupling}
\end{equation}%
where $C=k_{\parallel }R$, $e$ is electron charge, $\epsilon _{0}$ is the
vacuum electric constant. The results of coupling strength $g$ against the
nanowire radius $R$ and the distance between QD and nanowire edge $d$ with a
nanowire length of $L=10$ $\mu $m and a QD of exciton oscillator strength $%
f=100$ are shown in Fig. 2.

We have also carried out the detailed numerical simulations using the finite
element method (FEM) to verify the analytical results above. We use FEM
simulations to obtain the electromagnetic field solution and mode volume of
nanowire. In the inset of Fig. 3, we plot the mode profile ($m=0$) of
electric field ($|\epsilon _{i}E_{i}^{2}|$) with a nanowire of $R=70$ nm and
$L=10$ $\mu \text{m}$, where $i=1$, $2$ and $E_{1}$, $E_{2}$ denote the
electric field outside and inside the nanowire respectively. The
electromagnetic field presents the reflection and leakage at the end surface
of nanowire as shown in the inset, which are not involved in the analytical
model. The coupling strength of QD and nanowire interaction is given by $%
g=\hbar \lbrack \frac{1}{4\pi \epsilon _{1}\epsilon _{0}}\frac{\pi e^{2}f}{%
m_{e}\tilde{V}}]^{1/2}$ \cite{Claudio}, where $\tilde{V}$ is the
plasmon mode volume ($m=0$). The results of FEM simulations show
$g=0.41$ meV for a nanowire of $R=20$ nm and $d=30$ nm, while the
coupling constant is $g=0.49$ meV through Eq. (\ref{Coupling}). The
analytical results are larger than the results given by FEM
simulations because of the leakage of electromagnetic field at end
surface. To neglect the affect of the leakage of
electromagnetic field, we plot the normalized coupling strength $g_{N}=g(R)$/%
$g$($R=20$ nm, $d=0$) as a function of $R$ given by both FEM simulations and
the analytical derivation respectively in Fig. 3. It is found that the FEM
simulation results and analytical derivations agree closely.

\begin{figure}[h]
\includegraphics[width=0.7\columnwidth]{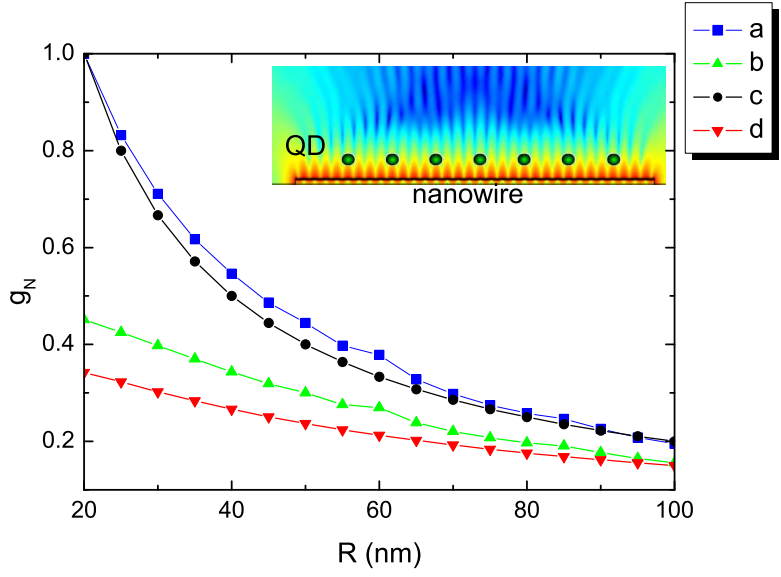}
\caption{ The normalized coupling strengths $g_{N}$ calculated using FEM
simulations are plotted as a function of $R$ at $d=0$ (line a), $d=30$ nm
(line b) with a nanowire length of $L = 10$ $\protect\mu$m. Compared with
the FEM simulations, the normalized coupling strengths $g_{N}$ obtained
using analytical model are plotted at $d=0$ (line c), $d=30$ nm (line d)
with a nanowire length of $L = 10$ $\protect\mu$m. Inset: the electric field
profile $|\protect\epsilon_i E_i|^{2}$ of the fundamental mode of a nanowire
with $R=70$ nm and $L=10$ $\protect\mu$m is calculated using FEM
simulations. }
\end{figure}

The dynamics of the coupled system of QDs and nanowire can be described by
the whole Hamiltonian:
\begin{equation}
H=H_{pl}+H_{QD}+H_{pl-QD}+H_{laser-QD}+H_{pl,\kappa }+H_{QD,\gamma }.
\end{equation}%
$H_{pl,\kappa }$ describes the plasmon decay through Landau damping, while $%
H_{QD,\gamma }$ describes the coupling of QD to Markovian reservoirs. We
consider several quantum dots coupled to a metallic nanowire and each dot is
irradiated with an identical frequency $\sigma ^{+}$ polarized laser light.
In the rotating frame at the laser frequency, the Hamiltonian is given by
\begin{equation}
H(t)=\sum_{j}[\delta _{L,j}|\uparrow \downarrow ,\Uparrow \rangle
_{j}\langle \uparrow \downarrow ,\Uparrow |+[(g_{j}a^{+}e^{i\Delta
_{j}t}+\Omega _{j}(t))|\uparrow \rangle _{j}\langle \uparrow \downarrow
,\Uparrow |+h.c.]],  \label{Hamiltonian}
\end{equation}%
where $\Delta _{j}=\delta _{L,j}-\delta _{pl,j}$.

To explore the applications of the plasmonics based quantum bus, we consider
the implementation of a controlled-phase (CPHASE) gate between two distant
spin qubits ($i$-th and $j$-th qubits). The adiabatic control is applied to
avoid the trion spontaneous emission and the plasmon decay. We use a laser
pulse with slowly changing Gaussian field amplitude $\Omega (t)=\Omega
_{0}e^{-(t^{2}/\tau ^{2})}$ and a constant detuning. The evolution operator
due to the adiabatic pulse is $U(t,t_{0})\equiv T\exp \left\{ -\frac{i}{%
\hbar }\int_{t_{0}}^{t}H(t^{\prime })dt^{\prime }\right\} $. Each phase
change of the four computational basis states $\{|00\rangle ,|01\rangle
,|10\rangle ,|11\rangle \}$ depends on adiabatic phase shift of single qubit
$\theta _{i}$ and nonlinear phase shift of both qubits $\phi _{ij}$. Assume
that the plasmon field is initially in the vacuum state. If the optical
pulse of amplitude is large enough to create a gate phase $\theta
_{i,j}=\phi _{00}-\phi _{01}-\phi _{10}+\phi _{11}=2$Re$\{\int dt\frac{%
\Omega _{i}(t)\Omega _{j}^{\ast }(t)g_{i}g_{j}^{\ast }}{\delta _{i}\delta
_{j}\Delta }\}=\pi $, a gate locally equivalent to the CPHASE gate is
achieved \cite{Zoller2001}. Up to a basis change of the target qubit and a
phase shift on the control qubit, CPHASE gate is equivalent to
controlled-not (CNOT) gate.

To estimate the performance of plasmonic quantum bus, we follow the standard
quantum theory of damping to calculate CPHASE gate fidelity in Markovian
approximation. The master equation for QDs and nanowire system can be
described by
\begin{equation}
{\frac{d\rho }{dt}}=-i[H_{s},\rho ]-{\frac{\kappa }{2}}(a^{\dagger }a\rho
+\rho a^{\dagger }a-2a\rho a^{\dagger })+\sum_{i}D[L_{i}]\rho ,  \label{ME}
\end{equation}%
where $\rho =\rho _{s}\otimes \rho _{pl}$ is density matrix of two qubits
and nanowire coupled system, $H_{s}=H_{pl}+H_{QD}+H_{pl-QD}+H_{laser-QD}$, $%
D[L_{i}]\rho =L_{i}\rho L_{i}^{\dagger }-\frac{1}{2}(L_{i}^{\dagger
}L_{i}\rho +\rho L_{i}^{\dagger }L_{i})$ and $L_{i}$ describes the trion
decay effect induced by various scattering channels such as phonon
environment and radiation field. For self-assembled InGaAs QDs, the key
decoherence parameter is trion radiative recombination time which is about $%
0.1$ ns. On the other hand, the effective lifetime of surface plasmon modes
is $(\Delta ^{2}/\Omega ^{2}(t))(\delta /g)^{2}/\kappa $, where $\kappa $ is
the decay rate of surface plasmon in nanowire. We consider two qubits
initialized into the $|\psi \rangle =(|00\rangle -|01\rangle -|10\rangle
+|11\rangle )/2$. Applying adiabatic Gaussian pulses with $\tau =\sqrt{\pi /2%
}\delta _{L,i}\delta _{L,j}^{\ast }\Delta /(\Omega _{0}^{2}g_{i}g_{j}^{\ast
} $) on both qubits, the output density matrix is $\rho $ following the
master equation (\ref{ME}) after the gate. The fidelity is defined as $%
\mathcal{F}=Tr [\sqrt{\sqrt{\rho^{\prime }}\rho\sqrt{\rho^{\prime }}}]$,
where the $\rho^{\prime }$ is the output density matrix following the master
equation without decay terms. We calculated numerically the fidelity of CPHASE gate of
two distant qubits versus the decay rate of the trion state $\Gamma$ and the quality factor of the plasmonic cavity $Q$ with optimum $\delta_L$ and $\Delta$ under the adiabatic approximation, as shown in Fig. 4. It is shown that, for $\Gamma=0.01$ ps$^{-1}$ and $Q=1000$, the fidelity of the two-qubit
gate can be $97\%$. The high-Q plasmonic microcavities ($Q > 1000$) have been reported in recent experiment \cite{Vahala}. Thus, high fidelity operations of two-qubit gate can be
achieved between two separated spin qubits coupled via quantum bus of
plasmon circuits.

\begin{figure}[tbp]
\includegraphics[width=0.7\columnwidth]{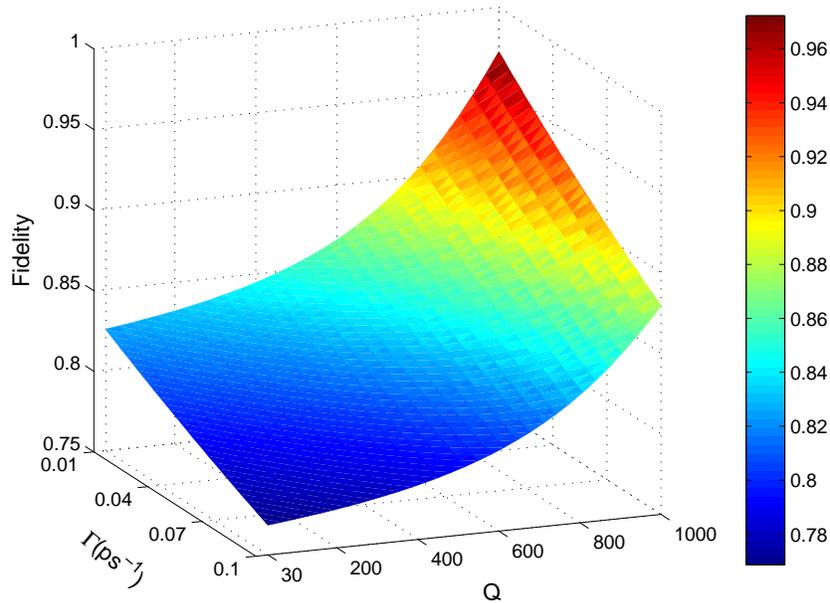}
\caption{Fidelity of CPHASE gate of two distant qubits as function of the decay rate of the trion state $\Gamma$ and the quality factor of the plasmonic cavity $Q$ with optimum $\delta_L$ and $\Delta$, calculated numerically under the adiabatic approximation.}
\end{figure}

We finally discuss potential realizations of distributed quantum computing
architectures which are based on plasmon mediated nonlocal spin-spin
interactions \cite{Hollenberg}. In the hybrid platform, each node which
consists of a nanowire and a few physical qubits, connects to a QD acting as
a transceiver qubit and all the transceiver qubits are coupled via a
plasmonic circuit or dielectric waveguide \cite{Lukin2007NP}. A key
technique to implement distributed quantum computing is performing a
nonlocal CNOT gate between qubit A in the $i$-th node and qubit B in the $j$%
-th node. We firstly apply polarized laser pulses on both $i$-th and $j$-th
transceiver qubits to prepare the qubits to be Einstein-Podolsky-Rosen state
$(\left\vert \downarrow \right\rangle _{i}\left\vert \uparrow \right\rangle
_{j}-\left\vert \uparrow \right\rangle _{i}\left\vert \downarrow
\right\rangle _{j})/\sqrt{2}$. After a set of sequential operations
including CNOT gate between qubit A and $i$-th transceiver qubit, CNOT gate
between qubit B and $j$-th transceiver qubit, single qubit measurement and
operation \cite{Eisert}, the nonlocal CNOT gate between qubit A and qubit B
is completed.

In summary, we develop a novel and scalable method to controllably couple
any distant electron spins in self-assembled quantum dots via surface
plasmon based quantum bus. A fully quantum mechanical approach is introduced
to describe the interaction between surface plasmon polariton of metal
nanowire and quantum dot spins. Virtual plasmon excitation is exploited to
overcome the decoherence of the system and a switchable long range
interaction is achieved between spin qubits. The proposed architecture is an
attractive approach for distributed quantum computation in a chip and
realizable by the present standard solid state chip technology.

\textbf{Acknowledgement} This work was supported by the National Basic
Research Program of China (Grants No. 2009CB929600, No. 2006CB921900), the
National Natural Science Foundation of China (Grants No. 10804104, No.
10874163, No. 10604052) and the State Key Program of the National Natural
Science Foundation of China (Grants No. 10934006).

\end{document}